\documentclass[aps,prc,twocolumn,showpacs,floatfix,nofootinbib,preprintnumbers,superscriptaddress,amsmath,amssymb]{revtex4-1}

\usepackage{graphicx}
\usepackage{dcolumn}
\usepackage{bm}
\usepackage{hyperref}
\usepackage[mathlines]{lineno}
\usepackage{subeqnarray}
\usepackage{cases}
\usepackage{times}
\usepackage{datetime}

\newcommand{\rr} {\boldsymbol{r}}

\begin{document}

\title{  Extension and parameterization of high-order density dependence in Skyrme forces }

\author{X.Y. Xiong}
\affiliation{State Key Laboratory of Nuclear
Physics and Technology, School of Physics, Peking University,  Beijing 100871, China}

\author{J.C. Pei}
\email{peij@pku.edu.cn}
\affiliation{State Key Laboratory of Nuclear
Physics and Technology, School of Physics, Peking University,  Beijing 100871, China}
\affiliation{State Key Laboratory of Theoretical Physics, Institute of Theoretical Physics, Chinese Academy of Sciences, Beijing 100190, China}

\author{W.J. Chen}
\affiliation{State Key Laboratory of Nuclear
Physics and Technology, School of Physics, Peking University,  Beijing 100871, China}
\begin{abstract}

The three-body force is indispensable in nuclear energy density functionals
which leads to a density dependent two-body term in the Hartree-Fock approach.
Usually a single factional power of density dependency has been adopted.
We consider the possibility of an additional higher-order density dependence in extended
Skyrme forces.
As a result, new extended Skyrme parametertizations based on the SLy4 force are obtained
and the improvements in descriptions of global nuclei have been demonstrated.
The higher-order term can also
 substantially affect nuclear properties in the high density region in general ways.

\end{abstract}



\maketitle

\section{introduction}

 The nuclear energy density functional theory (EDF) is a principal theoretical tool
 for descriptions of bulk properties and dynamics of the entire nuclear landscape.
 The main issues for nuclear EDF theory are concerns about its accuracy
and uncertainties. After early Skyrme Hartree-Fock calculations~\cite{vautherin}, the
Skyrme interactions~\cite{skyrme} have been widely used and have been demonstrated
to be very powerful~\cite{PGR}. In particular,
 the latest UNEDF Skyrme forces~\cite{markus1,markus2,markus3} based on large scale fitting procedures have remarkably improved the
accuracy. On the other hand, the limits of standard Skyrme forces have been reached
 and new ideas are desirable.

The Skyrme force is a very low-momentum (or soft) interaction with a
second-order momentum dependence and a three-body interaction that plays
an indispensable role in nuclear saturation~\cite{negele}.
 In \textit{ab initio} calculations,
 it is well known that the three-body force and many-body forces are generated due to suppressed
 (or renormalization cutoff) degrees of freedom of the two-body potential~\cite{grange}.
Following this picture, the remarkable three-body force contribution in soft Skyrme forces as
 well as Gogny forces~\cite{gogny} is then understandable.
 Recently the three-body force has become an attractive issue in \textit{ab initio} calculations~\cite{otsuka,hagen}
and Brueckner-Hartree-Fock calculations~\cite{zuo,zhang}.
In Hartree-Fock calculations, the three-body interaction
 can be considered as a density dependent two body term with $\rho^{\gamma}$, in which the
 density dependency power $\gamma$ should be 1.
However, the power $\gamma$ should be smaller than $2/3$ to get reasonable nuclear matter incompressibility~\cite{blaizot,sly4a}.
 In fact, a factional power of density dependency,  ranging from $1/6$ to 1, has been adopted in various Skyrme
 parameterizations~\cite{stone}. The parameterized density dependence is supposed to simulate complicated many-body correlations
 in addition to the three-body force and is still an open question.

 In the dilute Fermi systems of repulsive cores, the energy density functional can in principle be expanded
 in powers of $k_F$ (or $\rho^{1/3}$). The EDF expansion coefficients were calculated in a series of papers in 1957,
  by
 Huang-Yang~\cite{huang}, Lee-Yang~\cite{lee},
 Martin and De Dominicis~\cite{martin}, which takes the form,
 \begin{equation}
 \begin{array}{ll}
 \varepsilon = & \displaystyle \frac{3\hbar^2}{10m}(3\pi^2)^{2/3}\rho^{5/3}+\frac{\hbar^2\pi a}{m}\rho^2 \vspace{5pt}\\
 &\displaystyle +\frac{2\hbar^2a^23^{4/3}\pi^{2/3}}{35m}(11-2\ln2)\rho^{7/3} \vspace{5pt}\\
 & \displaystyle + 0.78\frac{\hbar^2a^33^{5/3}\pi^{10/3}}{2m}\rho^{8/3} \\
 \end{array}
 \label{leeyang}
 \end{equation}
 in which $a$ denotes a finite radius of particles.
 These high order terms reflect the increased kinetic energy by considering the finite radius of particles.
In the Local Density Approximation for dilute
 Fermi gases, its EDF form is the same as Skyrme EDF except for different density dependent terms~\cite{furnstahl}.
 The EDF of dilute Fermi gases
 involves terms with $\rho^{7/3}$, $\rho^{8/3}$, etc.
 While the standard Skyrme EDF involves a single density dependent term with $\rho^{2+\gamma}$.
 Actually, the $\rho^{7/3}$ term (corresponding to $\gamma$=1/3) can be mimicked by an effective three-body interaction~\cite{alex}.
  In contrast to the dilute Fermi systems,
 we note that generally the finite radii (volume effects) of nucleons have
 not been  explicitly considered, neither in realistic nor phenomenological nuclear forces.

 Our objective of this work is to consider the possibility of an additional higher order density dependence
 on top of the standard Skyrme force.
 In deeded, in the standard Skyrme force, a single density dependent
 term might be too simplistic. It is also of interdisciplinary interests to study the role of higher order
 density dependencies and the EDF connection between nuclei
 and dilute Fermi gases. We speculate that the high-order density dependencies
 can reflect short-range correlations beyond standard Skyrme forces, and impact in particular
the high-density behaviors and isovector properties.

The extensions of Skyrme forces have been developed by many efforts.
For example, the Brussels forces by considering the density dependence in
the momentum-dependent terms have been very successful~\cite{goriely}.
The effective pseudopotential of Skyrme-type EDF has been derived up to the sixth order~\cite{dobaczewski}.
 On the other hand, the connection between nuclear EDF parameters
 and \textit{ab initio} calculations has not yet been clear, although progresses have been achieved
 in the Density Matrix Expansion~\cite{stoitsov} and the Effective Field Theory~\cite{furnstahl}.

The higher-order density dependent terms have been considered in Refs.~\cite{cochet,agrawal06} in analogy to those for
the dilute Fermi gases, but with limited applications.
In this work, we are not intended to refit a fully new Skyrme parameter set from scratch but to improve the existed successful Skyrme parameters with
an additional higher-order density dependent term.
To this end, we concentrate on
 the momentum-independent terms, and manipulate the $t_0$, $t_3$ parameters and the extended term with $t_{3E}$.
Indeed, $t_0$, $t_3$ and $t_{3E}$ are directly entangled in the $s$-wave channel
 although they are also connected with other parameters.
For example, the correlations between parameters have been demonstrated by the Bayesian analysis~\cite{McDonnell}.
We choose to refit the momentum-independent terms based on the SLy4 force~\cite{sly4} and keep other parameters unchanged.
The SLy4 force is an ideal start because it has been widely used and has been the references for other parameterizations such as UNEDF forces.
Furthermore, SLy4 has a density dependence of $\gamma$=1/6 which is relatively small and may leave room for higher-order density dependencies.
Other successful Skyrme forces such as SkM*~\cite{skm} with $\gamma$=1/6 may also be improved with higher-order density dependencies.
Or say we consider an additional higher-order density dependent term for complementary to SLy4 in a perturbative manner.
It is turned out that the density dependent terms we employed are different from Refs.~\cite{cochet,agrawal06}.
We will then evaluate the performance of the extended new parameterizations compared to the original SLy4.

\section{Theoretical Methods}

We begin with the standard Skyrme interaction including the two-body and three-body terms~\cite{skyrme}, which takes the form,
\begin{eqnarray}
V_{\rm Skyrme}={\sum_{i<j}v_{ij}^{(2)}}+{\sum_{i<j<k}v_{ijk}^{(3)}}
\end{eqnarray}
\begin{equation}
\begin{array}{ll}
{v_{ij}^{(2)}}=&\displaystyle t_{0}(1+x_{0}P_{\sigma})\delta({\rr}_i-{\rr}_j)\vspace{5pt}\\
&+\displaystyle \frac{1}{2}t_{1}(1+x_1P_{\sigma})[\delta({\rr}_i-{\rr}_j){\bf{k}}^2+{\bf{k'}}^2\delta({\rr}_i-{\rr}_j)] \vspace{5pt}\\
&+\displaystyle t_2(1+x_2P_{\sigma}){\bf{k'}}\cdot\delta({\rr}_i-{\rr}_j){\bf{k}} \vspace{5pt}\\
&+\displaystyle iW_0(\sigma_i+\sigma_j)\cdot{\bf{k'}}\times\delta({\rr}_i-{\rr}_j){\bf{k}}
\end{array}
\label{skyrme}
\end{equation}
The three-body interaction in the Hartree-Fock calculations can be transformed
into a density dependent two-body interaction with $\rho^{\gamma}$. Correspondingly the power $\gamma$ should be 1 but usually it is
adjusted to be less than 1 in modern Skyrme forces to get reasonable incompressibilities.
\begin{equation}
\begin{array}{llll}
&\mathop{v_{ijk}^{(3)}}&=&t_{3}(1+x_3P_{\sigma})\delta({\rr}_i-{\rr}_j)\delta({\rr}_j-{\rr}_k) \vspace{5pt}\\
\Rightarrow & \mathop{v_{ij}^{(2)'}} &=& \displaystyle \frac{1}{6}t_3(1+x_3P_{\sigma})\rho({\bf{R}})^{\gamma}\delta({\rr}_i-{\rr}_j)\\
\end{array}
\label{skyrme3}
\end{equation}
In Eq.(\ref{skyrme}) and Eq.(\ref{skyrme3}) , $t_i$, $x_i$ and $W_0$ are the parameters of the Skyrme interaction,
and $\textbf{R}=(\rr_i+\rr_j)/2$.

In our extension,  by including an additional higher-order density dependent term,
the density dependent term is modified as,
\begin{equation}
\begin{array}{ll}
\mathop{v_{ij}^{(2)'}} = & \displaystyle \frac{1}{6}t_3(1+x_3P_{\sigma})\rho(\textbf{R})^{\gamma}\delta({\rr}_i-{\rr}_j)  \vspace{5pt} \\
& + \displaystyle \frac{1}{6}t_{3E}(1+x_{3E}P_{\sigma})\rho(\textbf{R})^{\gamma+\frac{1}{3}}\delta({\rr}_i-{\rr}_j)  \\
\end{array}
\label{extend}
\end{equation}
In this case, there will be 2 more additional parameters $t_{3E}$ and $x_{3E}$.
In the SLy4 force~\cite{sly4}, the power factor $\gamma$ takes $1/6$ and then we consider the next higher order power of $1/2=1/6+1/3$.
Actually such an extension is straightforward and can be easily implemented.
We noticed that similar extensions exactly according to the density dependent terms for dilute Fermi gases  have been discussed in the case of nuclear matter properties~\cite{cochet}.
  Later it has been investigated in finite nuclei but
negative coefficients are obtained for high-order terms~\cite{agrawal06}.
The density dependent terms we employed are different from~\cite{cochet,agrawal06}
 since our studies are based on SLy4, although we have similar motivations.
 It is useful to explore different combinations of density dependencies.
 Considering recent UNEDF forces with $\gamma$ around 1/3~\cite{markus1,markus2,markus3} that is
 between $1/6$ and $1/2$,  the two density dependent terms we employed should be reasonable.

Next we refit the extended Skyrme parameters for finite nuclei with the Simulated Annealing Method~\cite{sam}.
In this work, we only refit the momentum-independent parameters, $t_0$, $t_3$, $t_{3E}$, $x_0$, $x_3$, $x_{3E}$,
and keep others $t_1$, $t_2$, $x_1$, $x_2$, $W_0$ unchanged based on the SLy4 force.
In this way the influences of the extended higher-order density dependent term can be clearly illustrated, to avoid the
influences due to terms which involve momentum dependencies.

Our fitting procedure is similar to SLy4 as described in Ref.~\cite{sly4}. Briefly, we
minimize the quantity,
\begin{equation}
\begin{array}{ll}
\chi^2= & \displaystyle (\frac{e_{\infty}+16}{0.2})^2+  \displaystyle  \sum_{i}(\frac{E_{n(i)}-E_{UV14+UVII(i)}}{\Delta E_i})^2 \vspace{5pt}\\
&+ \displaystyle \sum_i (\frac{B_{t(i)}-B_{t(i)}^{\rm exp}}{2})^2+ \displaystyle \sum_i (\frac{R_{c(i)}-R_{c(i)}^{\rm exp}}{0.02})^2\\
\end{array}
\label{fit}
\end{equation}
where $e_{\infty}$ is the average energy per nucleon at the saturation point;  $E_{n(i)}$ and $E_{UV14+UVII(i)}$ are energies of the neutron matter from Skyrme forces and calculations of Wiringa et al.~\cite{wiringa} at different density points;
$B_t$ and $R_{c}$ denote total binding energies and charge radii of selected nuclei respectively.
In Eq.(\ref{fit}), the fitting uncertainties $\Delta E_{i}$ are not a constant but increase as densities increase~\cite{wiringa}.
The nuclear matter properties at the saturation point are important inputs for constraining parameters.
In the extended Skyrme functional, we have to slightly modify the equations of the pressure $P$, the incompressibility coefficient $K$,
the symmetry energy coefficient $a_s$.  The saturation density of 0.1595 fm$^{-3}$, the symmetry energy coefficient $a_s$ of 32 MeV, and the effective mass $m^{*}/m$ of 0.7 have not been adjusted~\cite{sly4}.
 Therefore, we adjust 6 parameters and have 4 free parameters $e_{\infty}$, $t_3$, $x_3$ and $x_{3E}$ to fit after
 taking into account the equations of nuclear matter properties. Then Skyrme parameters $t_0$, $x_0$, $t_{3E}$ can be determined through these expressions.

Our calculations are based on the axially-symmetric Hartree-Fock+BCS approach and the Simulated Annealing Method (SAM).
The Hartree-Fock+BCS equation is solved within axially-symmetric coordinate spaces using the \textsc{Skyax} solver~\cite{skyax}.
In this case deformed nuclei can be included in the fitting procedure compared to the original SLy4 force.
The calculations are performed in a 2D box of 30 fm and the uniform lattice space is taken as 0.5 fm.
The mixed pairing interaction~\cite{mix-pairing} has been adopted and the pairing strength is taken as $V_{p,n}$=500 MeV fm$^{-3}$ for protons and neutrons.
SAM~\cite{sam} is a general purpose algorithm for  multi-parameter fittings, based on
the Monte Carlo iterative solution method. SAM starts with a high temperature and randomly searches the minimum,
which accepts a worse solution than the current one with a certain probability.
As the temperature cools down, the searching space becomes smaller and the optimization is then realized.
The SAM has already been adopted in the fitting of full Skyrme parameters~\cite{agrawal}.

\begin{table*}
\caption{\label{table2}
Refits of SLy4 parameters with extend density dependent terms, compared to the original SLy4 force.
The different parameter sets are obtained by fitting different groups of nuclei, see text for details. We also refit
SLy4 without the extended term as labeled by SLy4$'$.
Other parameters in SLy4 not listed here are not adjusted.
The corresponding minimized $\chi^2$ of binding energies ($\chi^2_B$) and charge radii ($\chi^2_R$) of SLy4 are given in the brackets.  }
\begin{ruledtabular}
\begin{tabular*}{\textwidth}{@{\extracolsep{\fill}}lllllll}
 Parameters           &SLy4  &SLy4$'$({global})    &  Group1({magic})        &Group2({light})        &Group3({heavy})       &Group4({global})  \\
\hline

$ t_{0}(\rm{MeV\cdot fm^3})$            &-2488.91  & -2493.536  &-2132.57    &-2106.69     &-2310.79    &-2319.15 \\
$ t_{3}(\rm{MeV\cdot fm^{3(1+\frac{1}{6})}})$&13777.0  & 13809.324    &9366.12     &9036.30      &11549.45    &11660.70  \\
$ t_{3E}(\rm{MeV\cdot fm^{3(1+\frac{1}{2})}})$  &0  & 0  &2756.97     &2970.72      &1410.64     &1334.88  \\
$ x_0 $                                  &0.834  & 0.931  &0.968       &0.988        &0.972       &0.863  \\
$ x_3 $                                  &1.354  & 1.496  &1.922       &1.975       &1.715      &1.509 \\
$ x_{3E} $                               &0   &  0    &0.375       &0.530        &0.324       &0.622  \\
$ e_{\infty}$ (MeV)                            &-15.972         &-15.963         &-16.063        &-16.085  & -16.055      &-16.039  \\
$ K_{\infty}$ (MeV)                             &229.9         &230.31         &247.33       &248.84  & 239.19      &238.58  \\
$ \chi_R^2$                              &    &  10.35    &2.18(4.04)  &5.00(10.04)  &0.39(0.62)  &7.4(10.66)  \\
$ \chi_B^2$                              &     & 31.67    &0.83(2.46)  &3.83(6.82)   &19.26(78.28)&25.92(84.87)  \\
$ \chi_{\rm sum}^2$                          &   &  42.84      &3.67(6.91)  &9.79(17.306)&20.52(79.21)&33.82(95.94)  \\

\end{tabular*}
\end{ruledtabular}
\label{table1}
\end{table*}

\begin{table}
\caption{\label{table3}
Energies (in MeV) of neutron droplets in a $\hbar\omega$=10 MeV spherical trap calculated with SLy4 and the extended SLy4 (light) parameters,
 compared to \textit{ab-initio} results~\cite{potter}. The extended SLy4 parameters are given in Table \ref{table1}.}
     \begin{ruledtabular}                                                                                                                                                            \begin{tabular*}{\textwidth}{@{\extracolsep{\fill}}cccc}
~~N	&	SLy4	&  Extend (light) & \textit{ab initio}~~	\\
\hline
~~8	            &	120.81	&	124.03	   &	135.4(2)\footnote{No-core shell model calculations}~~	\\
~~10	            &	169.54	&	173.47	   &	187.1(6)$^{\textmd{a}}$~~	\\
~~12	            &	217.04	&	221.56	   &	237.1(8)$^{\textmd{a}}$~~	\\
~~14	            &	262.69	&	267.59	   &	286(1)$^{\textmd{a}}$~~	    \\
~~16	            &	309.28	&	314.28	   &	334(1)$^{\textmd{a}}$~~	    \\
~~18	            &	364.83	&	370.87	   &	386(3)$^{\textmd{a}}$~~	    \\
~~20	            &	418.56	&	425.50	   &	432.8(5)\footnote{Couple-cluster calculations}~~	\\
~~28	            &	656.18	&	663.84	   &	681(1)$^{\textmd{b}}$~~	    \\
~~40	            &	1061.23	&	1070.97	   &	1058(2)$^{\textmd{b}}$~~  	\\
~~50	            &	1420.19	&	1428.28	   &	1449(3)$^{\textmd{b}}$~~	    \\
\end{tabular*}
\end{ruledtabular}
\label{table2}
\end{table}

\begin{figure}[t]
  \includegraphics[width=0.5\textwidth]{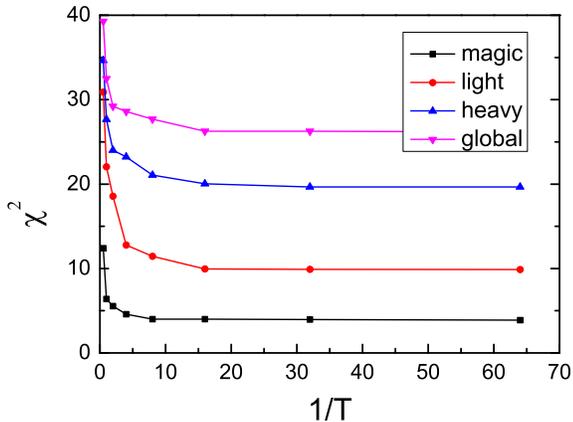}\\
  \caption{(Color online) The optimization of Skyrme parameter sets by adopting different fitting groups with the Simulated Annealing Method.
   The obtained $\chi^2$ values are displayed as a function of $1/T$, where $T$ denotes the temperature in SAM.  }
   \label{annealing}
\end{figure}

\begin{figure}[t]
  \includegraphics[width=0.5\textwidth]{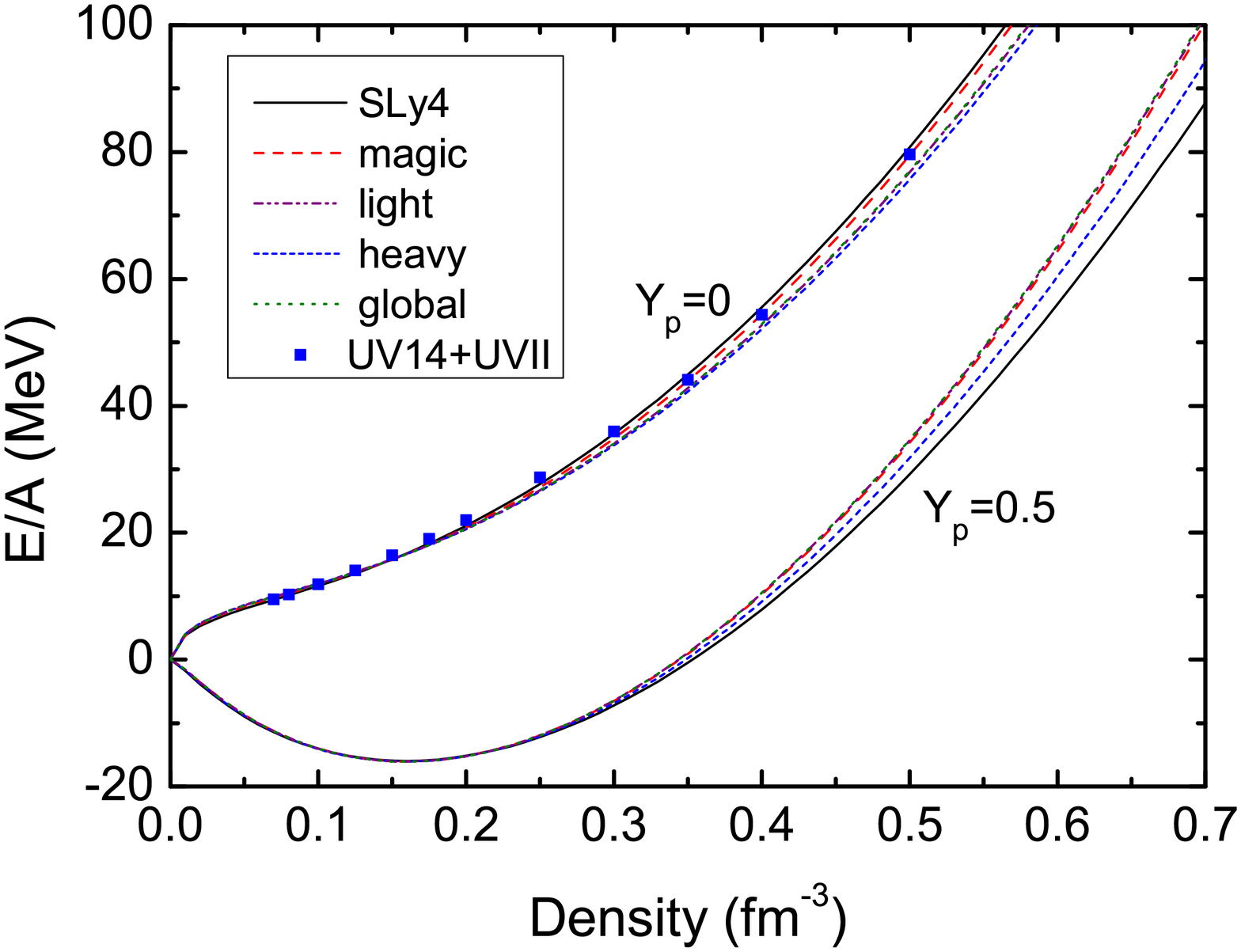}\\
  \caption{(Color online) Energies per particle in nuclear matter obtained
  by different Skyrme parameter sets (see Table \ref{table1}) as a function of density. $Y_p$=0 denotes for
  the pure neutron matter and $Y_p$=0.5 denotes for the symmetric nuclear matter. }
  \label{eos}
\end{figure}

 \begin{figure}[t]
   \includegraphics[width=0.49\textwidth]{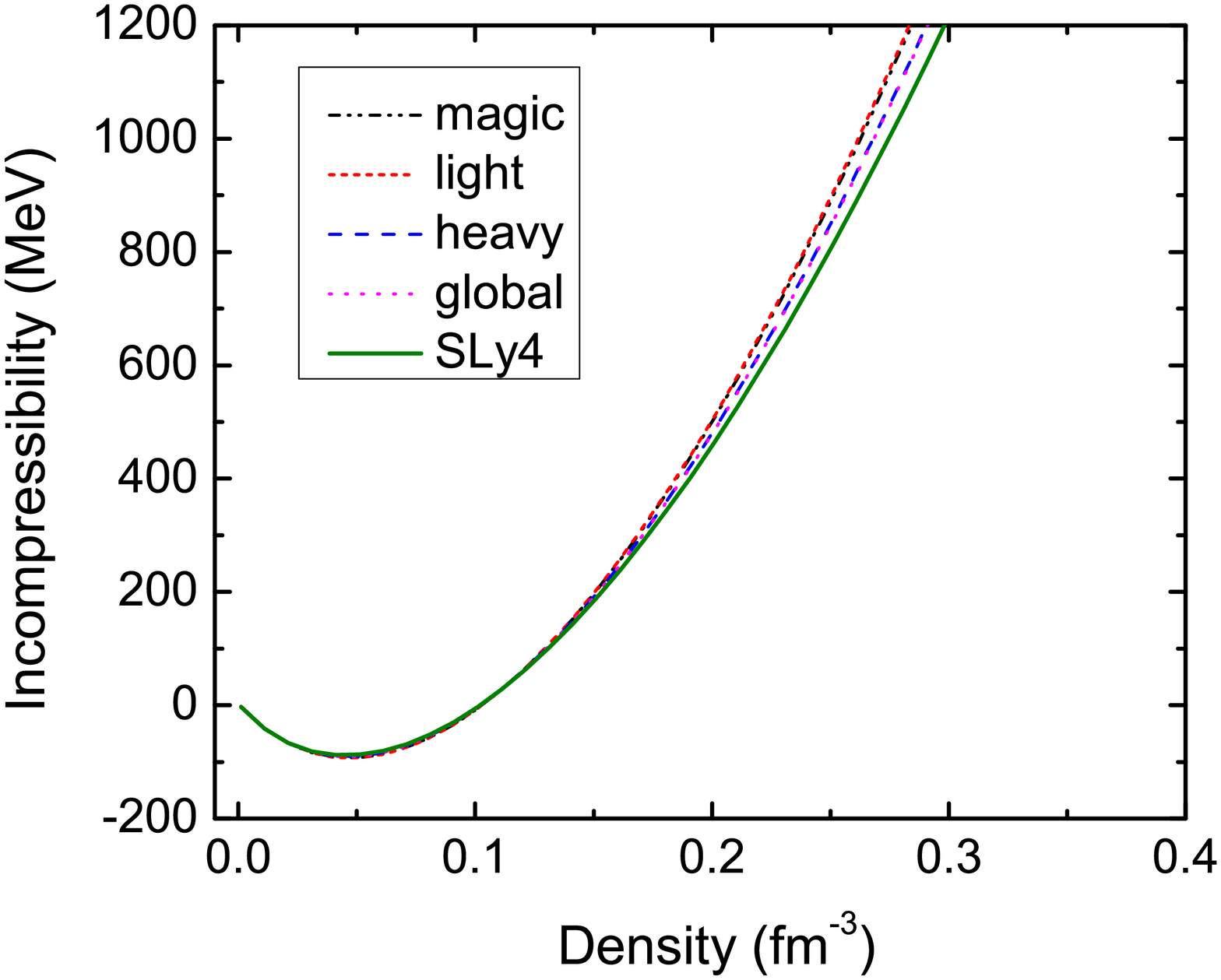}\\
   \caption{(Color online) Incompressibility coefficients $K$ as a function of nuclear matter density,
   corresponding to different Skyrme parameter sets in Table \ref{table1}.}
   \label{eosk}
 \end{figure}

\begin{figure}[t]
  \includegraphics[width=0.5\textwidth]{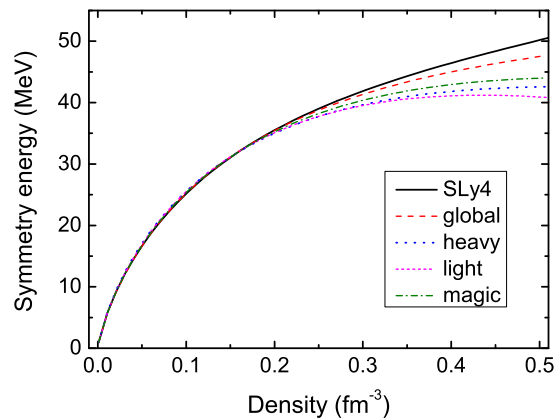}\\
  \caption{(Color online) Symmetry energies as a function of nuclear matter density,
   corresponding to different Skyrme parameter sets in Table \ref{table1}.}
   \label{eosas}
\end{figure}

\begin{figure}[t]
  \includegraphics[width=0.45\textwidth]{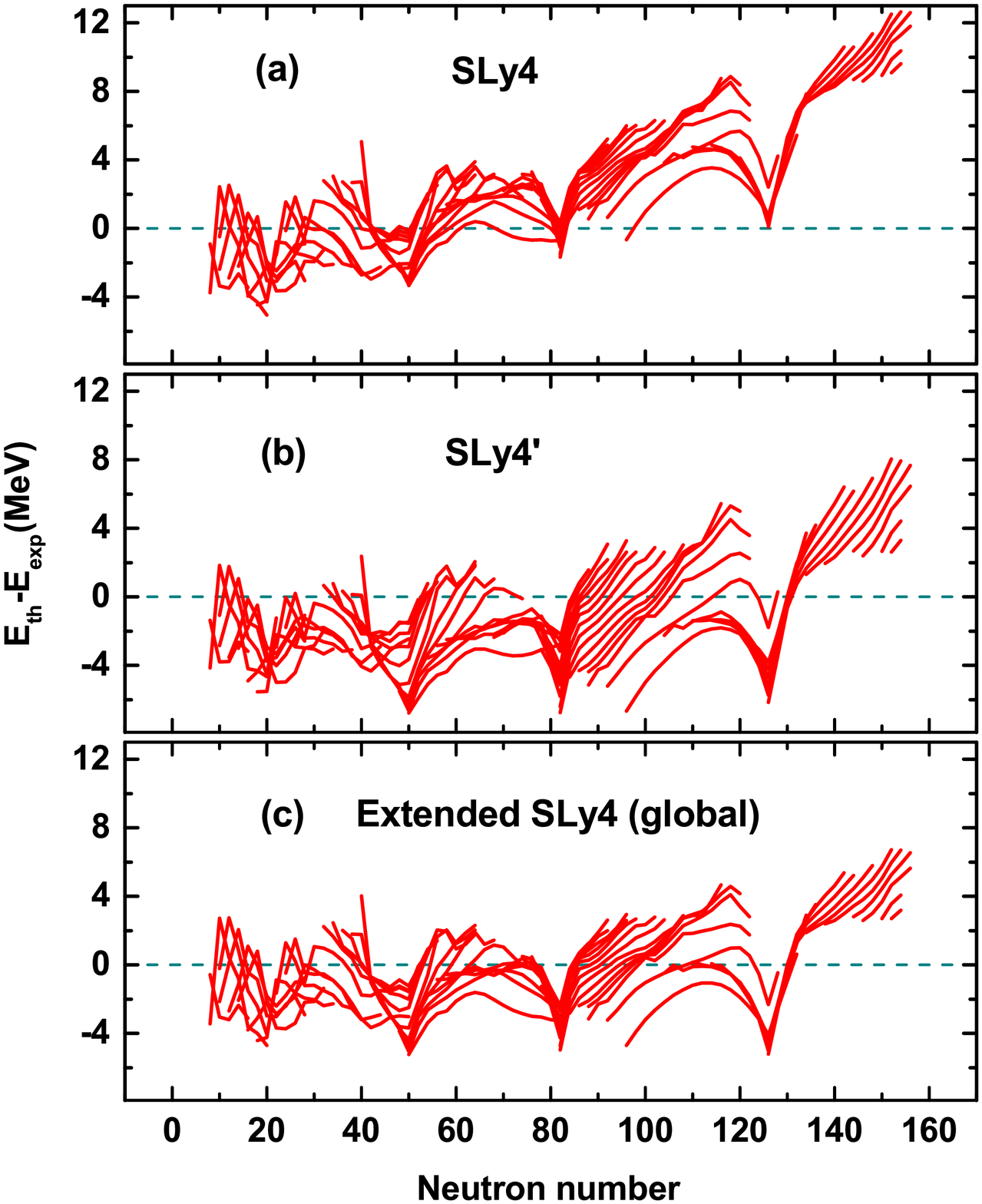}\\
  \caption{(Color online) Binding energy differences between theory and experiments~\cite{audi} for 603 even-even nuclei.
  The results are obtained by Hartree-Fock+BCS calculations with (a) the SLy4, (b) the refitted SLy4$'$  and (c) the extended SLy4
  (\rm{global}), respectively. See Table \ref{table1} for the parameter sets. }
  \label{global}
\end{figure}

  \section{\label{sec:result}Results and Discussions}

We refit the extended Skyrme force based on SLy4 with different groups of selected nuclei. In particular, we select 4 groups
aiming at different mass regions,

\textbf{Group1}: $^{40}\rm{Ca}$,$^{48}\rm{Ca}$,$^{56}\rm{Ni}$,
${}^{132}\rm{Sn}$, ${}^{208}\rm{Pb}$, for magic nuclei as adopted in the SLy4 fitting;

\textbf{Group2}: ${}^{16}\rm{O}$,${}^{36}\rm{Mg}$,${}^{40}\rm{Ca}$,${}^{48}\rm{Ca}$,${}^{50}\rm{Cr}$,${}^{56}\rm{Ni}$,${}^{78}\rm{Ni}$,${}^{100}\rm{Zr}$,
${}^{120}\rm{Sn}$,${}^{132}\rm{Sn}$, for light nuclei with $8\leqslant Z\leqslant50$;

\textbf{Group3}: ${}^{120}\rm{Sn}$,${}^{132}\rm{Sn}$,${}^{160}\rm{Gd}$,${}^{176}\rm{Hf}$,${}^{198}\rm{Pb}$,${}^{208}\rm{Pb}$,${}^{236}\rm{U}$,${}^{252}\rm{Fm}$,
${}^{260}\rm{Sg}$, for heavy nuclei with $Z\geqslant 50$;

\textbf{Group4}: ${}^{16}\rm{O}$,${}^{36}\rm{Mg}$,${}^{40}\rm{Ca}$,${}^{48}\rm{Ca}$,${}^{50}\rm{Cr}$,${}^{56}\rm{Ni}$,${}^{78}\rm{Ni}$,${}^{100}\rm{Zr}$,
${}^{120}\rm{Sn}$,${}^{132}\rm{Sn}$,${}^{160}\rm{Gd}$,${}^{176}\rm{Hf}$,${}^{198}\rm{Pb}$,${}^{208}\rm{Pb}$,${}^{236}\rm{U}$,${}^{252}\rm{Fm}$,
${}^{260}\rm{Sg}$, for the global mass region.

The experimental binding energies of these nuclei are taken from~\cite{audi}. The charge radii of 5 magic nuclei are taken from~\cite{sly4}.

Figure~\ref{annealing} illustrates the obtained $\chi^2$ of different SAM fittings as the temperature cools down. In the fitting procedure,
the initial temperature is 2 and decreases by a factor of 0.5 each step. The initial parameters are given as random values in
setted ranges. The initial ranges of $x_3$ and $x_{3E}$ are from $-2$ to 2. The range of $e_{\infty}$ is 16$\pm$0.15 MeV. The range of $t_3$ is from 8000 to 13777.
At the beginning, the $\chi^2$ is very large and becomes much smaller as the temperature lower than 0.125.  One full fitting procedure takes
about 200 iterations for the fitting of light nuclei and about 350 iterations for the global fitting, which takes 2$\thicksim$3 days.
After that, we polish these parameters by a second fitting with obtained parameters as initial values.

Table~\ref{table1} lists the obtained different parameterizations of extended SLy4 forces compared to the original SLy4 force.
Generally, the obtained $\chi^2$ values of charge radii and binding energies are reduced significantly, compared to the $\chi^2$ values of SLy4 shown in brackets.
 The incompressibility values $K_{\infty}$ and energies $e_{\infty}$ at the saturation density are also listed.
We can see that the existed parameters $t_0$, $t_3$ are reduced. The obtained $t_{3E}$ parameter of the higher-order term ranges from 1334.88 to 2970.72.
We noticed that in Refs.~\cite{colo,colo2}, with decreasing momentum cutoffs in
second-order Hartree-Fock calculations, the
adjusted parameters $t_0$, $t_3$ decrease and the density dependency power increases.
Our results with higher order terms are similar to that trend.
In the fittings of heavy and global nuclei, the parameters of the higher-order term are relatively small.
While in the fitting of light nuclei, the parameter of the higher-order term is the largest, which
is mainly responsible for improving the descriptions of charge radii.
This indicates that complex density dependencies are required for describing light nuclei.
In our tests, by reducing the fitting weights on charge radii,
the obtained higher-order term decreases.
In the heavy mass region,  the obtained new extended parameterizations still produce large deviations in binding energies.
We also refit the SLy4 with the global group, as labeled by SLy4$'$. The resulted $\chi^2$  is significantly reduced except for the descriptions of  charge radii.
This demonstrated that the extended higher-order density dependent term is important for descriptions of nuclear charge radii.

Figure~\ref{eos} displays the performances of extended parameterizations for energies of the neutron matter and the symmetric nuclear matter.
For the neutron matter, the energies given by UV14+UVII~\cite{wiringa} are also shown.
Since our fittings have included the UV14+UVII results, we see energies are close to the original SLy4.
In the case of neutron matter, energies of all the extended parameterizations are slightly decreased in the high density region
compared to the original SLy4.
While in the case of symmetric nuclear matter, energies are slightly increased.
Note that large uncertainties in the high density region have been taken into account in the fitting procedures.
The energies of neutron droplets in a $\hbar\omega$=10 MeV trap have also been calculated, as displayed in Table \ref{table2}.
The \textit{ab initio} calculations were done by the no-core shell model and the couple-cluster model with the
Chiral two-body (N3LO) plus three-body (N2LO) forces~\cite{potter}.  Generally, the SLy4 force underestimates
the energies compared to the Chiral force. While the extended SLy4 force can bring more binding energies. This
is consistent with the situation of neutron matter energies as shown in Fig.\ref{eos}.

To study the behaviors of extended SLy4 forces for nuclear matter properties, the incompressibility
and the symmetry energy are displayed in Fig~\ref{eosk} and Fig.\ref{eosas} respectively.
In Fig.\ref{eosk}, the extended forces all generate larger incompressibility values at high densities compared to the SLy4 force.
The parameters for light nuclei with the largest higher order term produce the highest incompressibility.
This causes very little influences in the low density region. It is understandable that the increased
incompressibility is due to the repulsive higher-order density dependence.
In Fig.\ref{eosas}, the symmetry energies as a function of densities are displayed. It can be seen that
the symmetry energies decrease in the high density region by considering the higher order density dependent term.
Again the parameters for light nuclei produce the softest symmetry energy.
There has been some favorable arguments about the soft symmetry energy~\cite{chen}, in contrast to the observation
of a neutron star with two solar mass. While heavy ion collision experiments can not give clear constraints about the
symmetry energy in dense nuclear matter~\cite{chen}.  In fact, by studying the expressions of nuclear matter properties,
we note that the behaviors of the extended higher-order density dependence shown in Figs.~(\ref{eosk}-\ref{eosas}) are general.
We conclude
that the inclusion of a higher-order density dependent term would impact the
equation of state and isovector properties in the high density region (supra-saturation density), which are important
 for studying neutron star structures.

Finally, the global calculations of binding energies of 603 even-even nuclei are shown in Fig.\ref{global}.
The calculations are done with the axially-symmetric Hartree-Fock+BCS solver \textsc{Skyax}, and with the original SLy4, the refitted SLy4$'$ and the extended SLy4 (global) parameter sets.
As we can see that the results of the extended parameterization are similar to the SLy4$'$ results.
The original SLy4 remarkably underestimates the binding energies of heavy and superheavy nuclei.
The extended SLy4 and SLy4$'$ have improved the global descriptions by including some heavy nuclei in the fitting procedure.
For all the nuclei, the resulted root-mean-square (rms) deviations of the extended SLy4, SLy4$'$ and SLy4 are 2.30 MeV, 2.94 MeV and 4.37 MeV, respectively.
In the region of $Z\leqslant82$, the rms deviations of the extended SLy4, SLy4$'$ and SLy4 are 2.00 MeV, 2.71 MeV, 3.18 MeV, respectively.
In the region of $Z\leqslant50$, the rms deviations of the extended SLy4, SLy4$'$ and SLy4 are 2.15 MeV, 2.88 MeV, 1.97 MeV, respectively.
Therefore we can say that the extended SLy4 force has globally improved the descriptions of binding energies, compared to the refitted SLy4$'$ and the original SLy4.
Note that the extended global parameterization set has a relatively small higher-order density dependent term, as displayed in Table \ref{table1}.
We see that SLy4, SLy4$'$ and the extended SLy4 are not able to describe the binding energies of magic nuclei and deformed nuclei simultaneously.
Or say the shell effects are overestimated as they are based on a small effective mass of 0.7~\cite{sly4}.
Presently the effective mass and the related $t_1$, $t_2$ parameters have not been adjusted in the
extended SLy4 forces.
Indeed, UNEDF Skyrme forces with large effective masses around 1 are very successful in describing  binding energies of the whole nuclear landscape~\cite{markus1}.
On the other hand, the Brussels Skyrme force with an effective mass of 0.8 and additional collective corrections for deformations also works very successful~\cite{goriely13}.
In addition, SLy4 is unable to describe the surface tension compared to calculations with microscopic center-of-mass corrections~\cite{bender}.
Actually, our main goal in this work is to study the influences of the extended higher-order density
dependence,  by refitting only the momentum-independent parameters. The global performance of the extended SLy4 is very encouraging.
To further improve the overall performance, the adjustments of all the Skyrme parameters including the extended term by choosing properly physics inputs will be our next step.

\section{summary}

In summary, we studied the extension of Skyrme forces
by including a higher-order density dependent term, according to
our speculation that a single density dependent term might be
too simplistic.
This is in analogy to high order density dependent terms in the EDF for dilute Fermi gases.
We studied the influences by adjusting only the momentum-independent parameters
based on the SLy4 force, with two density dependent terms of $\rho^{1/6}$ and $\rho^{1/2}$.
The obtained new extended parameterizations indicate that the strengthes
of the higher order term $\rho^{1/2}$ are dependent on different fitting regions.
We have seen that
 the higher-order term can improve descriptions of binding energies of global nuclei and neutron droplets as expected.
 We also noticed that the higher-order term is important for descriptions of charge radii.
The extended parameterization set obtained by fitting light nuclei has
the largest higher order term, indicating complex density dependencies are required for light nuclei.
Furthermore, we demonstrated that the extended density dependence can impact nuclear matter properties
 particularly in the supra-saturation density region with general behaviors, although the extrapolations of Skyrme forces
to the high density region still have large uncertainties.
In conclusion, our studies have provided some insights and opportunities for future developments of Skyrme energy density functionals.

\begin{acknowledgments}
 Useful suggestions by J. Vary, F.R. Xu, W. Nazarewicz, N. Van Giai are gratefully acknowledged.
 This work was supported by the National Natural Science Foundation of China under Grants No.11375016, 11522538, 11235001,
and by the Research Fund for
the Doctoral Program of Higher Education of China (Grant
No. 20130001110001);and the Open Project Program of State Key Laboratory of Theoretical Physics,
Institute of Theoretical Physics, Chinese Academy of Sciences, China (Grant No. Y4KF041CJ1).
We also acknowledge that some computations in this work were performed in the Tianhe-1A supercomputer
located in the Chinese National Supercomputer Center in Tianjin.
\end{acknowledgments}


\end{document}